\documentclass[aps,prb,twocolumn,showpacs,showkeys,preprintnumbers,superscriptaddress,amsmath,amssymb,]{revtex4-1}
\usepackage{graphicx}
\usepackage{dcolumn}
\usepackage{bm}
\usepackage{hyperref}
\usepackage[dvips]{color}
\usepackage{multirow}
\begin{document}
\title{Structures, optical properties, and electrical transport processes of SnO$_2$ films with oxygen deficiencies}
\author{Yu-Chen Ji}
\author{Hua-Xing Zhang}
\affiliation{Tianjin Key Laboratory of Low Dimensional Materials Physics and
Preparing Technology, Department of Physics, Tianjin University, Tianjin 300072,
China}
\author{Xing-Hua Zhang}
\affiliation{Hebei University of Technology, School of Material Science and and Engineering, Tianjin 300130, China.}
\author{Zhi-Qing Li}
\email[Author to whom correspondence should be addressed. Electronic mail: ] {zhiqingli@tju.edu.cn}
\affiliation{Tianjin Key Laboratory of Low Dimensional Materials Physics and
Preparing Technology, Department of Physics, Tianjin University, Tianjin 300072,
China}
\date{\today}
\begin{abstract}
The structures, optical and electrical transport properties of SnO$_2$ films, fabricated by rf sputtering method at different oxygen partial pressures, were systematically
investigated. It has been found that preferred growth orientation of SnO$_2$ film is strongly related to the oxygen partial pressure during deposition, which provides an effective way to tune the surface texture of SnO$_2$ film. All films reveal relatively high transparency in the visible range, and both the transmittance and optical band gap increase with increasing oxygen partial pressure. The temperature dependence of resisitivities was measured from 380 K down to liquid helium temperatures. At temperature above $\sim$$80$ K, besides the nearest-neighbor-hopping process, thermal activation processes related to two donor levels ($\sim$$30$ and $\sim$100 meV below the conduction band minimum) of oxygen vacancies are responsible for the charge transport properties.  Below $\sim$$80$ K, Mott variable-range-hopping
conduction process governs the charge transport properties at higher temperatures, while Efros-Shklovskii variable-range-hopping conduction process dominates the transport properties at lower temperatures. Distinct crossover from Mott type to Efros-Shklovskii type variable-range-hopping conduction process at several to a few tens kelvin are observed for all SnO$_2$ films.
\end{abstract}
\maketitle
\section{Introduction}
In recent years, great interest has been focused on SnO$_{2}$ due to its importance in many technical fields, such as gas sensors,\cite{Batzill} solar cells,\cite{Ren} and optoelectronic devices.\cite{Pan} The nominally undoped (unintentional doped) SnO$_{2}$ is a transparent metal-oxide semiconductor which is natively n-type and usually has a resistivity in the range of 1-100 $\Omega$ cm.\cite{Bansal} The optical transparency of SnO$_{2}$ thin films can be up to 97\% in the visible range (for films of thickness 0.1-1 $\mu$m) with a direct optical band gap of 3.6-4.2 eV.\cite{Kilic,T.Serin} The conductivity of SnO$_{2}$ film can be greatly enhanced by doping small amount of fluorine, antimony, tantalum and other dopants.\cite{Stjerna,Li,Lee} Currently, fluorine doped SnO$_{2}$ (FTO) film is one of the widely used transparent conducting oxide (TCO) materials. In addition, the p-type conductivity, which is crucial in fabricating transparent p-n junction, was recently observed in Li, In, Al, and N doped SnO$_{2}$,\cite{Pan} which further stimulates the studies on SnO$_{2}$ materials. The influences of the fabricating method and fabricating condition on the structures, electrical transport and optical properties of SnO$_2$ films had been extensively investigated.\cite{Stjerna} The electrical transport properties of different forms of SnO$_{2}$ such as nanobelt, nanowire, and thin films had also been studied.\cite{Viana,Bazargan,Ma} However, the origins of electrical conductivity as well as the charge transport mechanisms of SnO$_2$ have not been fully understood, which seriously limits optimizing the properties of this material. Thus detailed measurements of the temperature behavior of resistivities over a wide temperature range and then extracting the charge transport mechanisms in SnO$_{2}$ films are necessary and nontrivial.
\par
In the present paper, we systematically investigated the temperature dependence of resistivities from liquid helium temperatures to 380 K for seven polycrystalline SnO$_{2}$  films deposited by rf sputtering method at different oxygen partial pressures. (It is known that the carrier concentrations in SnO$_{2}$ films are seriously influenced by the oxygen contents.)  The electrical transport process was found to be dominated by thermal activation and variable-range-hopping (VRH) processes at high and low temperatures, respectively. Particularly, a transition from Mott type to Efros-Shklovskii (ES) type VRH conduction process was observed for all the samples. The variations in surface morphologies and optical properties with oxygen partial pressure for the films were also  studied.
\section{Experimental method}
The SnO$_{2}$ thin films were deposited on quartz glass by standard rf sputtering method. A SnO$_{2}$ target (with diameter of 60 mm and 5 mm thick), synthesized by the traditional ceramic process, was used as the sputtering source. SnO$_{2}$ powders with 99.9\% purity were ground by ball milling for 12 h before being compressed into a disk. The disk was calcined at 1673 K in air for 12 h and then furnace cooled. The base pressure of the chamber was pumped to $1.0$$\times$$10^{-4}$ Pa and then a mixture gas of Ar and O$_{2}$ (both were 99.999\% purity) was introduced as the sputtering gas. Oxygen partial pressure ($O_{pp}$) was controlled (0, 0.5\%, 1\%, 1.5\%, 5\%, 10\%, and 20\%) by adjusting the flux ratio of Ar to O$_{2}$ and the pressure of chamber was maintained at 1 Pa. The substrate temperature and the sputtering power were kept at 973 K and 150 W, respectively.
\par
Film thickness ($\sim$$1$ $\mu$m) was determined by a surface profiler (Dektak, 6 m). Crystal structures of the films  were measured in a powder x-ray diffractometer (D/MAX-2500, Rigaku) with Cu $K\alpha$ radiation. The surface morphology of SnO$_{2}$ thin films was characterized by a scanning electron microscope (SEM, S-4800, Hitachi). The resistivity was measured by the standard four-probe technique. A Keithley 236 unit and a Keithley 2182A nano-voltmeter were used as the current source and voltmeter, respectively. The temperature environment was provided by a physical property
measurement system (PPMS-6000, Quantum Design). Optical absorption and transmittance spectra were measured in a UV-VIS-NIR scanning spectrophotometer (UV-3101 PC, SHIMADZU).
\begin{figure}[htp]
\begin{center}
\includegraphics[scale=0.7]{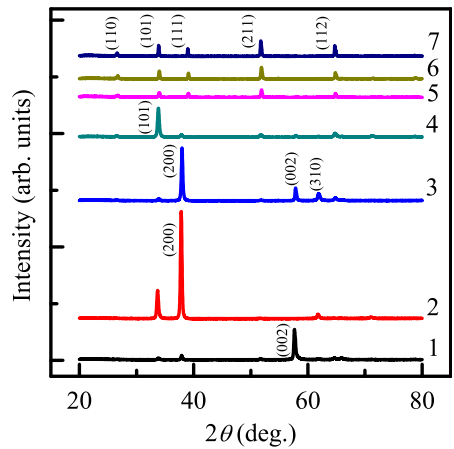}
\caption{(color online) XRD patterns of SnO$_{2}$ thin films deposited under different oxygen partial pressures (0, 0.5\%, 1\%, 1.5\%, 5\%, 10\%, and 20\%).}\label{FIG1-XRD}
\end{center}
\end{figure}
\section{Results and discussion}
Figure~\ref{FIG1-XRD} shows the XRD patterns of the SnO$_{2}$ thin films deposited at different oxygen partial pressures. Samples were numbered from 1 to 7 according to the oxygen partial pressures (0, 0.5\%, 1\%, 1.5\%, 5\%, 10\%, and 20\%) during deposition. Comparing to the standard PDF card (No. 41-1445), one can know that the films have tetragonal rutile-type structure and there is no secondary phase. The lattice constants are $a$$\approx$4.742 \r{A}, $c$$\approx$3.189 \r{A}, and insensitive to the oxygen contents, while the preferred growth orientation varies with the variation in oxygen partial pressure. The strongest peak is from the diffraction of (002) plane for sample No. 1, while it becomes (200) plane for sample Nos. 2 and 3. For sample No. 4, the diffraction corresponding to the (101) plane is the strongest peak. With further increasing oxygen contents, the relative intensity of (101) peak is obviously reduced and the relative intensity of (111), (211), and (112) peaks increased sharply. The different preferred growth orientations would lead to different surface texture of the films, which is important in improving the efficiency of solar cells.\cite{Ren} Our results indicate that altering the oxygen partial pressure in the sputtering process is an effective way to tune the texture of SnO$_2$ films.
\begin{figure}[htp]
\begin{center}
\includegraphics[scale=0.7]{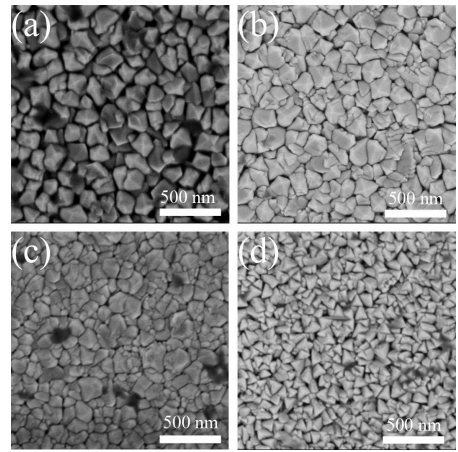}
\caption{(color online) SEM images of SnO$_{2}$ thin films deposited under oxygen partial pressures of: (a) 0, (b) 1\%, (c) 1.5\%, (d) 5\%.}\label{FIG2-SEM}
\end{center}
\end{figure}
\par
Figure~\ref{FIG2-SEM} shows the SEM images for four representative films. The variation in the texture of the films can also be seen in the morphologies of the films. Films Nos. 1 to 3 show square based pyramid-like structures. Film No. 4 shows a gravel-like morphology. Films Nos. 5 to 7 exhibit triangular pyramidal shape grains. According to the morphology simulation results by Smith,\cite{Smith} the variation in surface morphology is consistent with the change in the preferred growth orientation observed in XRD measurements. The mean grain size $D$ was obtained by surveying $\sim$60 grains in each film and was listed in Table~\ref{Table-Thermal}. The grain size shows a decrease trend with increasing oxygen partial pressure, which may be related to the variation in the preferred orientations of the films.
\begin{figure}[hbp]
\begin{center}
\includegraphics[scale=0.7]{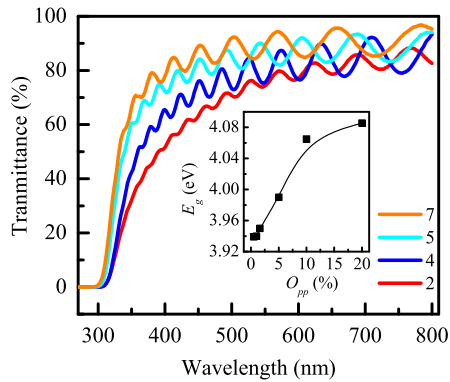}
\caption{(color online) Room temperature optical transmission spectra of four SnO$_{2}$ thin films, as indicated. The inset shows the values of optical band gap versus oxygen partial pressure.}\label{FIG3-T}
\end{center}
\end{figure}
\par
Figure~\ref{FIG3-T} shows the optical transmittance ($T_R$) versus wavelength for four representative SnO$_{2}$ films, as indicated. Here the transmittance $T_R$ is the intensity ratio of transmitted light to incident light, i.e., $T_R$$=$$I/I_0$. From Fig.~\ref{FIG3-T}, we can see that all films exhibit relatively high transparencies in the visible range (390-780 nm). At a certain wave range of visible spectra, the transparency of the film enhances with increasing oxygen partial pressure. In addition, the absorption edge slightly shifts to shorter wavelength side with increasing oxygen content. Since the relation between transmittance and absorption coefficient ($\alpha$) is  $T_R$$=$$\exp(-\alpha d)$ (where $d$ is the thickness of the film), one can readily obtain the absorption coefficient versus photon energy $h\nu$ from the data in Fig.~\ref{FIG3-T}. Theoretically, the relationship between absorption coefficient and the incident photon energy can be written as\cite{Pankove}
\begin{equation}\label{Eq.(Eg)}
(\alpha h \nu)^{1/n}=A(h\nu-E_{g}),
\end{equation}
where $A$ is a constant and $E_{g}$  is the band gap. Using $n$$=$$1/2$  for direct band gap transition, the values of $E_{g}$ were obtained by extrapolating the linear portion of the plot to $(\alpha h \nu)^{2}$$=$$0$.  The inset of Fig.~\ref{FIG3-T} shows the obtained  $E_g$ value versus oxygen partial pressure. Clearly, the optical band gap of the film increases with oxygen partial pressure which is in accordance with the blue shift of the absorption edge. The oxygen vacancies are often treated as the main reason for the opaque of SnO$_{2}$ thin films.\cite{Choi} Thus, the enhancement in the transmittance and optical band gap of the films can be attributed to the reduction of oxygen vacancies with increasing oxygen partial pressure.
\par
\begin{figure}[htp]
\begin{center}
\includegraphics[scale=0.7]{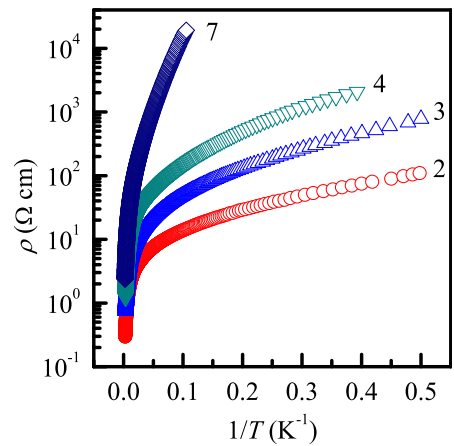}
\caption{(color online) Variation in the logarithm of resistivity with reciprocal temperature for four SnO$_{2}$ thin films, as indicated.}\label{FIG4-rhoT}
\end{center}
\end{figure}

Figure~\ref{FIG4-rhoT} shows the resistivities as functions of reciprocal temperature from 380 K down to liquid helium temperatures for four representative films, as indicated. Inspection of this figure indicates that the resistivity increases with decreasing temperature over the entire measured temperature range which shows a typical semiconductor behavior in the electrical transport property. At a certain temperature, the film deposited under lower oxygen partial pressure possesses smaller resistivity, e.g., the resistivity of sample No. 7 is about 1000 times larger than that of sample No. 1 at 10 K. Our observation supports K{\i}l{\i}\c{c} and Zunger's prediction that oxygen vacancies are the main electron suppliers in SnO$_{2}$.\cite{Kilic}
\begin{table}[htp]
\caption{\label{Table-Thermal} Values of relevant parameters for seven oxygen deficient SnO$_{2}$ polycrystalline films. $\rho$$_{i}$ ($i$$=$1, 2, 3) and $E_{i}$ are defined in Eq.~(\ref{Eq.(Arrhenius)}). $D$ is the mean grain size of the films.}
\begin{ruledtabular}
\begin{center}
\begin{tabular}{ccccccccc}
Film& $O_{pp}$ & $D$ & $\rho_{1}$ & $E_{1}$ & $\rho_{2}$ & $E_{2}$ & $\rho_{3}$ & $E_{3}$\\
 No. & (\%) & (nm) & ($\Omega$ cm) & (meV) & ($\Omega$ cm) & (meV) & ($\Omega$ cm) & (meV)\\\hline
1&	0&	    141&   0.067&	105&	0.297&	30.8&	1.01&	3.82\\
2&	0.5&	144&   0.055&	91&	    0.228&	32.6&	1.38&	4.12\\
3&	1&	    132&   0.064&	114&	0.316&	34.5&	3.06&	4.45\\
4&	1.5&	98.3&  0.078&	102&	0.621&	35.7&	5.59&	5.60\\
5&	5&	    78.1&  0.127&	108&	2.92&	31.3&	16.9&	8.29\\
6&	10&	    70.3&  0.097&	119&	1.42&	36.4&	18.6&	6.39\\
7&	20&	    75.4&  0.154&	114&	1.77&	37.0&	17.7&	7.81\\
\end{tabular}
\end{center}
\end{ruledtabular}
\end{table}
\par
\begin{figure}[hbp]
\begin{center}
\includegraphics[scale=0.7]{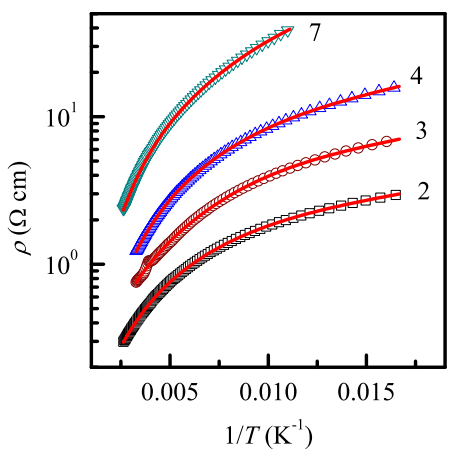}
\caption{(color online) Logarithm of resistivity as a function of reciprocal temperature for four SnO$_{2}$ thin films from $\sim$80 to $\sim$380 K, as indicated. The symbols are the experimental data and the solid curves are least-squares fits to Eq.~(\ref{Eq.(Arrhenius)}).}\label{FIG5-Thermal}
\end{center}
\end{figure}

We analyze the carrier transport processes in high temperature regions. In Fig.~\ref{FIG5-Thermal}, we replotted the resistivities as functions of reciprocal temperature from $\sim$80 to $\sim$380 K for the four representative SnO$_{2}$ thin films. The high temperature conduction process in n-type semiconductor usually comes from the electrons hopping from the shallow donor levels to the conduction band which can be expressed by the Arrhenius function $\rho(T)^{-1}$$=$$\rho^{-1}_{0}e^{-E_{a}/k_{B}T}$. However, no straight lines can be found in Fig.~\ref{FIG5-Thermal}. According to Samson and Fonstad,\cite{Samson} the oxygen vacancies in SnO$_2$ can form two donor levels, one is a shallow donor level ($\sim$30 meV below the conduction band minimum), the other is an intermediately deep donor level ($\sim$100 meV below the conduction band minimum). In addition, the electrons in the impurity band can also take part in the conducting process by hopping from an occupied site to the nearest empty site via phonon assistance [which is called nearest-neighbor-hopping (NNH) process] at intermediate temperature range.\cite{Lien} (Note that the NNH conduction also has an Arrhenus form.) Thus, we tentatively fit our $\rho(T)$ data by
\begin{equation}\label{Eq.(Arrhenius)}
\rho(T)^{-1}=\rho^{-1}_{1}e^{-E_{1}/k_{B}T}+\rho^{-1}_{2}e^{-E_{2}/k_{B}T}+\rho^{-1}_{3}e^{-E_{3}/k_{B}T},
\end{equation}
where $\rho_1$, $\rho_2$, and $\rho_3$ are prefactors independent of temperature, $E_1$ and $E_2$ are the activation energies
associated with the two donor levels, and $E_3$ is the activation energy of the NNH process.
The theoretical predications of Eq.~(\ref{Eq.(Arrhenius)}) are also shown in Fig.~\ref{FIG5-Thermal} (the solid curves), from which one can see that
the measured  $\rho(T)$ data can be well described by Eq.~(\ref{Eq.(Arrhenius)}). [Precisely, samples Nos. 1 to 4 can be fitted by Eq.~(\ref{Eq.(Arrhenius)}) from 60 to 380 K, and samples Nos. 5 to 7 can be fitted from 90 to 380 K.] The relevant parameters obtained in the fitting processes are listed in Table~\ref{Table-Thermal}. Inspection of Table~\ref{Table-Thermal} indicates that the values of $E_1$ and $E_2$ are $\sim$$100$$\pm$20 meV and $\sim$$30$$\pm$7 meV, respectively, which are consistent with the activation  energies of the two donor levels of oxygen vacancies.\cite{Samson} While the value of $E_3$ varies from 3.82 to 8.29 meV,
which is in good accordance with the activation energy of NNH conduction.\cite{Bansal,Viana,N.Serin}
\begin{table*}
\caption{\label{Table-VRH} Values of relevant parameters for seven oxygen deficient SnO$_{2}$ polycrystalline films studied in this work. $\rho$$_{M}$, $T_{M}$ and $\rho$$_{ES}$, $T_{ES}$ are defined in Eq.~(\ref{Eq.(rho-Mott)}) and Eq.~(\ref{Eq.(rho-ES)}), respectively. $T_{\text{cross}}$ and $T'_{\text{cross}}$ are the crossover temperatures from Mott to ES VRH conduction which were obtained from theoretical calculation and observation of the data plot, respectively. $\Delta_{CG}$ is the value of the Coulomb gap. $\overline{W}_{\text{hop,Mott}}$ and $\overline{W}_{\text{hop},ES}$ are the average hopping energies of electrons.}
\begin{ruledtabular}
\begin{center}
\begin{tabular}{cccccccccc}
Film No.& $\rho_{M}$ & $T_{M}$ & $\rho_{ES}$ & $T_{ES}$ & $T_{\text{cross}}$ & $T'_{\text{cross}}$ & $\Delta_{CG}$ & $\overline{W}_{\text{hop,Mott}}$ & $\overline{W}_{\text{hop},ES}$  \\
 & ($\Omega$ cm) & (K) & ($\Omega$ cm) & (K) & (K) & (K) & (meV) & (meV) & (meV)\\\hline
1&		0.151&	2780&	1.86&	31.6&	5.76&	4&	0.262&  1.57$\times10^{-4}$$T^{3/4}$&	2.43$\times10^{-4}T^{1/2}$\\
2&		0.302&	2159&	2.95&	26.2&	5.07&	6&  0.223&  1.47$\times10^{-4}$$T^{3/4}$&	2.21$\times10^{-4}T^{1/2}$\\
3&	    0.282&	7172&	5.65&	49.6&	5.49&	8&	0.320&	1.98$\times10^{-4}$$T^{3/4}$&	3.04$\times10^{-4}T^{1/2}$\\
4&	 	0.361&	13053&	8.84&	81.2&	8.07&	8&  0.496&	2.30$\times10^{-4}$$T^{3/4}$&	3.88$\times10^{-4}T^{1/2}$\\
5&      0.060&	155608&	2.55&	688&	48.7&	40&	 3.55&	4.28$\times10^{-4}$$T^{3/4}$&	1.13$\times10^{-3}T^{1/2}$\\
6&	    0.039&	187461&	2.04&	746&	47.5&	40&	 3.65&	4.49$\times10^{-4}$$T^{3/4}$&	1.18$\times10^{-3}T^{1/2}$\\
7&	    0.043&	195370&	2.27&	771&	48.6&	40&	 3.75&	4.53$\times10^{-4}$$T^{3/4}$&	1.20$\times10^{-3}T^{1/2}$\\
\end{tabular}
\end{center}
\end{ruledtabular}
\end{table*}
\par
The empty sites among the nearest neighbors will decrease dramatically with decreasing temperature. At sufficient low temperatures, the NNH process would be quenched and the electrons will hop to empty sites that are not nearest to them but have smaller energy discrepancies.\cite{Gantmakher} In this case, there are usually two forms of hopping. At relatively high temperatures, Mott suggested the VRH conduction in the form of\cite{Davis,Gantmakher,Shklovskii}
\begin{equation}\label{Eq.(rho-Mott)}
\rho(T)=\rho_{M}e^{(T_{M}/T)^{1/4}},
\end{equation}
for three dimensions, where $\rho_{M}$  is a temperature independent prefactor and $T_{M}$ is a characteristic temperature. The Mott VRH conduction theory is based on a constant density of states (DOS) near the Fermi level. The characteristic temperature can be written as\cite{Shklovskii}
\begin{equation}\label{Eq.(T-Mott)}
T_{M}=\frac{18}{k_{B}N(E_{F})\xi^{3}},
\end{equation}
where $N(E_{F})$  is the DOS without long range coulomb interactions and $\xi$  is the relevant electron localization length. When the temperature is further decreased, the Coulomb interactions between the charge carriers should be considered. In this situation, ES
found that the electronic DOS in the vicinity of $E_F$ is
no longer a constant, and a soft gap (Coulomb gap) will form in the DOS near the Fermi energy. When the influence of the Coulomb gap can no longer be ignored, the temperature dependence of resistivity would become\cite{Shklovskii,Efros}
\begin{equation}\label{Eq.(rho-ES)}
\rho(T)=\rho_{ES}e^{(T_{ES}/T)^{1/2}},
\end{equation}
where $\rho_{ES}$ is a temperature independent prefactor and  $T_{ES}$ is a characteristic temperature. The characteristic temperature  $T_{ES}$ is given by\cite{Shklovskii,Efros}
\begin{equation}\label{Eq.(T-ES)}
T_{ES}=\frac{\beta_{1}e^{2}}{\epsilon{\xi}k_{B}},
\end{equation}
where $\beta_{1}$  is a constant with a value of $\simeq$2.8, $\epsilon$ is the dielectric constant of the material. In ES VRH model, the width of the Coulomb gap is\cite{Efros}
\begin{equation}\label{Eq.(w-gap)}
\Delta_{CG}=\frac{e^3\sqrt{N(E_F)}}{\epsilon^{3/2}}.
\end{equation}
\begin{figure}[htp]
\begin{center}
\includegraphics[scale=0.9]{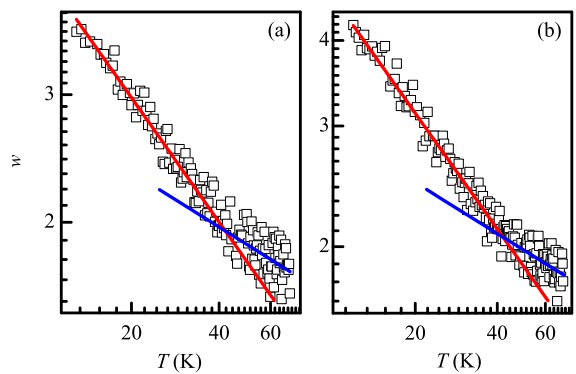}
\caption{(color online) Log-log plot of $w$$=$$-\partial\ln \rho/\partial\ln T$  versus $T$  for samples (a) No. 5 and (b) No. 7. The solid lines are the linear fits to the data.}\label{FIG6-wT}
\end{center}
\end{figure}
\par
To check whether the Mott and ES types VRH processes exist in our samples, we plotted $w$ against $T$ on double logarithmic scales, where $w$ is defined as $w(T)$$=$$-\partial\ln\rho/\partial\ln{T}$.\cite{Hill} Using the general form of VRH conduction $\rho(T)$$=$$\rho_{0}\exp(T_{0}/T)^{x}$, one can obtain the expression $\log w$$\simeq$$\log(xT^{x}_{0})$$-$$x\log T$. Thus when the hopping mechanisms dominate the carrier transport processes, one can get the magnitudes of exponent $x$ from the slopes of the linear parts in $\log w$ versus $\log T$ data. Figure~\ref{FIG6-wT} shows  $w$ as a function of $T$ in double-logarithmic scales from $\sim$10 to 70 K for two representative films (Nos. 5 and 7). Clearly, there are two distinct linear parts in each curve, which suggests the existence of two different hopping mechanisms in our films at low temperatures. The values of $x$ obtained from each plot are $\sim$0.56 and  $\sim$0.25 at lower temperature region and higher temperature region, respectively, which suggests that the crossover from ES VRH ($x$$\sim$$0.56$) to Mott VRH ($x$$\sim$0.25) process occurs with increasing temperature. For samples Nos. 5 and 7,  both the crossover temperatures are around 40 K.
\par
\begin{figure}[ftp]
\begin{center}
\includegraphics[scale=0.9]{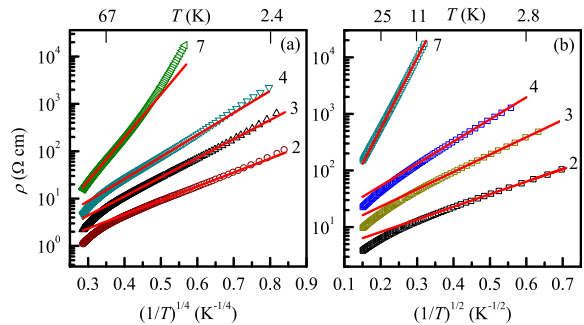}
\caption{(color online) (a) Logarithm of resistivity as a function of $T^{-1/4}$ for four SnO$_{2}$ thin films, as indicated. The symbols are the experimental data and the straight solid lines are least-squares fits to Eq.~(\ref{Eq.(rho-Mott)}), (b) Logarithm of resistivity as a function of $T^{-1/2}$ for four SnO$_{2}$ thin films, as indicated. The symbols are the experimental data and the straight solid lines are least-squares fits to Eq.~(\ref{Eq.(rho-ES)}).}\label{FIG7-VRH}
\end{center}
\end{figure}

The experimental $\rho(T)$ data at higher and lower temperature regions are least-squares fitted to Eq.~(\ref{Eq.(rho-Mott)}) and Eq.~(\ref{Eq.(rho-ES)}), respectively, and the results are shown in Fig.~\ref{FIG7-VRH}. For samples Nos. 1 to 4, the experimental data can be well described by Eq.~(\ref{Eq.(rho-Mott)}) in temperature range of $\sim$10 K to $\sim$40 K. For samples Nos. 5 to 7, the corresponding temperature range is $\sim$40 K to $\sim$70 K. Outside the above temperature regions, the $\rho(T)$ data deviates from the predication of Eq.~(\ref{Eq.(rho-Mott)}). The deviation parts are quite coincide with the predication of Eq.~(\ref{Eq.(rho-ES)}) below $\sim$10 K for samples Nos. 1 to 4 and $\sim$30 K for samples Nos. 5 to 7. Thus, the crossover from the Mott to ES VRH occurs near $\sim$10 K for samples Nos. 1 to 4 and $\sim$40 K for samples Nos. 5 to 7. The fitting parameters $\rho_M$, $T_M$,  $\rho_{ES}$, and $T_{ES}$ are listed in Table~\ref{Table-VRH}.
\par
In the Mott VRH and ES VRH theories, the average hopping energies of electrons are given by\cite{Davis,Shklovskii}
\begin{equation}\label{Eq.(W-Mott)}
\overline{W}_{\text{hop,Mott}}=\frac{1}{4}k_{B}T\left(\frac{T_{M}}{T}\right)^{1/4}
\end{equation}
and
\begin{equation}\label{Eq.(W-ES)}
\overline{W}_{\text{hop},ES}=\frac{1}{2}k_{B}T\left(\frac{T_{ES}}{T}\right)^{1/2},
\end{equation}
respectively. The hopping energy in Eq.~(\ref{Eq.(W-Mott)}) should be equal to that in  Eq.~(\ref{Eq.(W-ES)}) at $T_{\text{cross}}$, thus the exact value of $T_{\text{cross}}$ can be written as
\begin{equation}\label{Eq.(T-cross)}
T_{\text{cross}}=\frac{16T^{2}_{ES}}{T_{M}}.
\end{equation}
The values of the crossover temperature $T_{\text{cross}}$ together with that estimated from $\log w$-$\log T$ plots (written as $T'_{\text{cross}}$) are listed in Table~\ref{Table-VRH}. Since the transition
from the Mott VRH conduction behavior to the ES VRH
conduction behavior is a smooth process, rather than a
sharp crossover, the slight discrepancy between $T_{\text{cross}}$ and $T'_{\text{cross}}$ is physically reasonable.
\par
We further analyze the applicability of the Mott and ES VRH conduction laws to our films.
Combining Eqs.~(\ref{Eq.(T-Mott)}), (\ref{Eq.(T-ES)}), and (\ref{Eq.(w-gap)}), the width of the Coulomb gap can be rewritten as
\begin{equation}\label{Eq.(CG)}
\Delta_{CG}=0.9k_{B}\left(\frac{T^{3}_{ES}}{T_{M}}\right)^{1/2}.
\end{equation}
Thus the values of $\Delta_{CG}$ are obtained and also listed in Table II, from which one can see that $\Delta_{CG}$ increases with increasing oxygen partial pressure. The average hopping energies $\overline{W}_{\text{hop,Mott}}$ and $\overline{W}_{\text{hop},ES}$ are calculated using Eq.~(\ref{Eq.(W-Mott)}) and Eq.~(\ref{Eq.(W-ES)}), respectively. The criterion of $\overline{W}_{\text{hop,Mott}}$$\geqslant$$2\Delta_{CG}$ in the Mott VRH theory\cite{Rosenbaum} requires a temperature  higher than $\sim$5 K, $\sim$7 K, and $\sim$40 K for samples Nos. 1 to 3, No. 4, and Nos. 5 to 7, respectively. Taking account of the crossover temperatures listed in Table~\ref{Table-VRH}, we can see that the criterion of the Mott VRH conduction is fully satisfied for all samples. However, the requirement of $\overline{W}_{\text{hop},ES}$$<$$\Delta_{CG}$ in the ES VRH theory is hardly satisfied for all samples, since it requires temperatures to be lower than $\sim$2~K and $\sim$10~K for samples Nos. 1 to 4 and Nos. 5 to 7, respectively. The discrepancy between experimental results and theoretical predictions in the $\overline{W}_{\text{hop},ES}$ was also observed in other systems,\cite{Lien,Rosenbaum,Yildiz} which still needs further investigation.
\section{conclusion}
In summary, we systematically investigated the charge transport processes of SnO$_2$ thin films, which were deposited by rf sputtering method at different oxygen partial pressures, from 380 K down to liquid helium temperatures. The influences of oxygen partial pressure on the structures and optical properties were also studied. The preferred crystal growth  direction as well as the surface morphology of the films can be tuned by varying oxygen partial pressure in the deposition process.
All films reveal relatively high transparency in the visible range, and both the transmittance and optical band gap increase with increasing oxygen partial pressure.
In the temperature range $80$ K$\lesssim$$T$$\lesssim$$380$ K, the NNH conduction process together with two thermal activation conduction processes, one from the shallow donor levels and the other from the intermediately deep donor levels, are the main charge transport mechanisms. Below $\sim$80 K, the Mott type and ES type VRH sequentially governs the conduction processes with decreasing temperature. Our work establishes a quite complete picture of the overall
electrical conduction mechanisms in the SnO$_2$ polycrystalline films from liquid-helium temperatures up to 380 K.

\begin{acknowledgments}
We would like to thank Z. M. Lu for his contribution in SEM measurements. This work was supported by the NSF of China through grant no. 11174216, Research Fund for the Doctoral Program of Higher Education, and Tianjin City NSF through grant no. 10JCYBJC02400.
\end{acknowledgments}

\end{document}